\documentclass[
]{ceurart}

\usepackage{xspace}
\newcommand{\R}{\mathbb{R}}
\newcommand{\tarski}{\textsc{Tarski}\xspace}
\newcommand{\zthree}{\textsc{Z3}\xspace}
\newcommand{\cvc}{\textsc{cvc5}\xspace}
\usepackage{booktabs}

\usepackage{subcaption}
\begin{document}
\copyrightyear{2024}
\copyrightclause{Copyright for this paper by its authors.
  Use permitted under Creative Commons License Attribution 4.0 International (CC BY 4.0).}

\conference{9th International Workshop on Satisfiability Checking and Symbolic Computation, July 2, 2024, Nancy, France, Collocated with IJCAR 2024}

\title{Symbolic Computation for All the Fun}


\author[1]{Chad E. Brown}
\author[1]{Mikol\'a\v s Janota}[%
orcid=0000-0003-3487-784X,
email=mikolas.janota@gmail.com,
url=http://people.ciirc.cvut.cz/~janotmik/%
]
\author[2]{Mirek Ol\v s\'ak}[%
orcid=0000-0002-9361-1921%
]
\address[1]{Czech Technical University in Prague, CIIRC, Czechia}
\address[2]{University of Cambridge}

\begin{abstract}
    Motivated by the recent 10 million dollar AIMO challenge, this
    paper targets the problem of finding all functions conforming to a given
    specification. This is a popular problem at mathematical competitions and it
    brings about a number of challenges, primarily, synthesizing the possible
    solutions and proving that no other solutions exist. Often, there are
    infinitely many solutions and then the set of solutions has to be captured
    symbolically. We propose an approach to solving this problem and evaluate
    it on a set of problems that appeared in mathematical competitions and
    olympics.
\end{abstract}

\begin{keywords}
  AIMO \sep%
  synthesis \sep%
  quantifier elimination \sep%
  SMT
\end{keywords}

\maketitle%

\section{Introduction}\label{sec:intro}

The \emph{AIMO challenge}\footnote{\url{https://aimoprize.com/}} promises to
award \$10 million to an AI model that will be able to win a gold medal at the
\emph{International Mathematical Olympiad
(IMO)}\hspace{1pt}\footnote{\url{https://www.imo-official.org/}}. In this paper we draw
attention to the fact that symbolic computation tools can give a significant
boost to a system aiming at this challenge. We target a popular problem
encountered in mathematical competitions, which is finding \emph{all} functions
adhering to a certain specification. The general task appears near
impossible. Indeed, obtaining a single function is already difficult and typically achieved by
synthesis approaches, such as SyGuS~\cite{sygus}, or up by specific procedures for
dedicated fragments of logic, cf.~\cite{hozzova-cade23,hozzova-ijcar24,ratschan-mfcs23,bradley-vmcai06,ge-moura-cav09}.  However, the
problems in the competitions are constructed so that they have ``nice'' solutions.
As a motivational example, consider the problem that asks us to find all
functions $f$ from $\R$ to $\R$ satisfying the following identity.

\begin{equation}
    \forall xy:\R.f(x+y) = xf(y) + yf(x)
\end{equation}

Substituting $y$ with $0$ gives $\forall x:\R.f(x)=f(0)x$, which tells us that $f$ must be
linear and that it is fully determined by the value of $f(0)$. Further, substituting
$x$ with $0$ shows that $f$ has to be $0$ everywhere, i.e., the solution is
$f(x)= 0$ (technically, $f=\lambda x.\,0$). A solution is expected to give a clear description of how all
possible $f$ are calculated, possibly parametrized by constants. To verify the
correctness of the solution, one needs to prove the following biconditional.

\begin{equation}
    (\forall xy:\R.f(x+y) = xf(y) + yf(x)) \Leftrightarrow (\forall x:\R.f(x)=0)
\end{equation}

Note that the problem of finding all solutions is not fully formal in the problem statement,
that is, we expect a ``reasonable description'' of the set of all the solutions. For example,
the following would \emph{not} be considered a valid answer: The solution consists of all the functions
satisfying $f(x+y) = xf(y) + yf(x)$ for all $x,y\in\mathbb{R}$.

A problem of such a form ``find all solutions satisfying $X$'' sparked some debate within
the community about how they should be formally handled.
How exactly we handle such questions is explained in Section~\ref{sec:solving}.



\section{Problems Description}\label{sec:problems}
As a source for our problems we have used a collection
compiled by V{\'\i}t Musil~\cite{musil} comprising problems from several
sources, including international competition and some (easier) problems from the 
Prague Seminar in Mathematics\footnote{\url{https://prase.cz/info/info_en.php?lang=1}}.
We have transcribed the problems into SMT2, where each original problem is divided into 3 types of queries.

\begin{itemize}
  \item The satisfiability of the specification if there is a solution (problems \texttt{find});
      for the example above:
    \[ \forall xy:\R.f(x+y) = xf(y) + yf(x) \]

  \item The unsatisfiability of the specification together with the negation of all proposed solutions
      (problems \texttt{prove}); for the example above:
    \[ \forall xy:\R.f(x+y) = xf(y) + yf(x) \land \exists x:\R. f(x)\neq 0 \]
  \item The check that every proposed solution is indeed a solution to the specifications
      (problems~\texttt{check}),
    i.e., the unsatisfiability of the proposed solution together with the negation of the specification;
    for the example above:
    \footnote{In most cases, the SMT solver can get rid of $f$ completely by macro detection.}
    \[ \exists xy:\R.f(x+y) \neq xf(y) + yf(x) \land \forall x:\R. f(x)=0 \]
\end{itemize}

This way, we are asking SMT solvers to solve particular steps, although these steps do not necessarily cover
the full solution of a functional equation. We are not providing a formal description of what a valid solution
is allowed to consist of (that could be further work using for example SyGuS~\cite{sygus}). Rather, we use our own method for finding
the set of all solutions of a particular form, described in Section~\ref{sec:solving}.
The SMT2 problems are made available as a GitHub repository\footnote{\url{https://github.com/MikolasJanota/FuncProbs}}
describing 87 problems. In some cases, the \texttt{check} query is split into multiple SMT2 files, if there are multiple distinct solutions.

We remark that during this work we uncovered several bugs in our translation to
SMT from the source material. But also, we have found an issue in the original
source material. Namely the problem C9 (Cvi\v cen\'\i~9) was unsatisfiable in
the original~\cite{musil} and we created two versions of the problem as two
different ways of correcting it (\texttt{C9} and \texttt{C9a}).







\subsection{Solutions of Selected Problems}

Section~\ref{sec:solving} describes our approach to solving the problem by
trying to prove that it fits a fixed template, in particular, we focus on
functions that can be expressed as polynomials. In most cases, this is probably
\textbf{not} how a human would solve it.
It is the subject of future work to make this connection.
For comparison, we show here three example problems together with their original (human) solutions:
\begin{itemize}
\item Problem U24, which is among the most interesting ones that we were able to fully solve;
\item Problem C12, which we were not able to solve but it does not require any advanced technique, so it could be feasible to solve it;
\item Problem U2, which showcases a problem requiring induction to be solved, and we assume is currently out of reach.
\end{itemize}

\paragraph{Example Problem U24 (Baltic Way 1998-7):}
Find all functions $f : \mathbb{R} \to \mathbb{R}$ such that for any pair of real numbers $x,y$, the following identity holds
\[ f(x) + f(y) = f(f(x)f(y)). \]

\paragraph{Solution:} First, we notice that the image of $f$ is closed under addition:
if $a,b$ are the values of $f$ at points $x$, $y$, then also $a+b$ is a value of $f$,
in particular at point $f(x)f(y) = ab$.
Fix arbitrary $x_1$, and denote $a = f(x_1)$.
By the previous observation, we can find also $x_2$ and $x_4$ such that $f(x_2) = 2a$,
and $f(x_4) = 4a$.

Now, we plug into the functional equation the assignments $(x = x_2, y = x_2)$, and $(x = x_1, y = x_4)$.
\begin{align*}
  4a = f(x_2) + f(x_2) &= f(f(x_2)f(x_2)) = f(4a^2)\\
  5a = f(x_1) + f(x_4) &= f(f(x_1)f(x_4)) = f(4a^2)
\end{align*}
We conclude $4a = 5a$, and consequently $a = 0$. Since we started with arbitrary $f(x) = a$, the function must be constant zero, which indeed satisfies the equation.
For this problem, the SMT solver \cvc was able to demonstrate that $f$ must be
constantly $0$ by the enumerative strategy~\cite{janota2021fair}.
The proof found is analogous to the human proof.\footnote{%
To obtain instantiations from \cvc, use \texttt{--no-e-matching --enum-inst --dump-instantiations --produce-proof}.}
It does \textbf{not} contain the general observation that the domain is closed under addition,
which is possible because this property is only needed for the points appearing in the proof.
The proof is also far less legible because it does not denote subexpressions by new constants (such as $a=f(x_1)$).

\bigskip

\paragraph{Example Problem C12 (IMO 1992-2):}
Find all functions $f : \mathbb{R} \to \mathbb{R}$ such that for any pair of real numbers $x,y$, the following identity holds
\[ f(x^2+f(y)) = y+f(x)^2. \]
\paragraph{Solution:} Setting $x=0$ gives $f(f(y)) = y + f(0)^2$. The right-hand side $y + f(0)^2$ is a bijection, so also the left-hand side $f\circ f$ is a bijection $\mathbb R\to\mathbb{R}$, and that can only happen when $f$ itself is a bijection.

Since $f$ is a bijection, we can consider an argument $b$ such that $f(b) = 0$. Let us substitute $(x=b, y=0)$ and $(x=-b, y=0)$ to the original equation.
\[
0 = 0 + f(b)^2 = f(b^2 + f(0)) = f((-b)^2 + f(0)) = 0 + f(-b)^2 = f(-b)^2.
\]
We obtained $f(-b) = 0$. By injectivity, $-b = b$, so $b = 0$, and $f(0) = 0$.
This also simplifies our first observation to $f(f(y)) = y$.

Now we prove that the function is increasing. Any pair of numbers $a < b$ can be expressed as
$a = f(y), b = x^2 + f(y)$ for some $x,y$, $x\neq 0$. Combining the original equation with the same equation having substituted $x = 0$, we get
\[
f(a) = f(f(y)) = y < y + {f(x)}^2 = f(x^2 + f(y)) = f(b),
\]
concluding that $f$ is increasing.

Consider any $y$ where $y \leq f(y)$. Since $f$ is increasing, also $f(y) \le f(f(y)) = y$.
Analogously, if $f(y) \leq y$, we obtain $y = f(f(y)) \le f(y)$. Thus the statements $y \leq f(y)$ and
$f(y) \leq y$ are equivalent for any $y$, leaving the only option $f(y) = y$,
which is a function satisfying the given equation.

\bigskip

\paragraph{Example Problem U2 (Cauchy equation):}
Find all \textbf{increasing} functions $f : \mathbb{R} \to \mathbb{R}$ such that for any pair of real numbers $x,y$, the following identity holds
\[ f(x) + f(y) = f(x+y). \]

\paragraph{Proof sketch:} By setting $y = 0,x,2x,3x,\ldots$, we obtain
by induction that $f(nx) = nf(x)$ for every non-negative integer $n$, in particular $f(0) = 0$,
and if we fix $x=1$ and denote $f(1) = c$, we obtain the following equation for all non-negative integers $x$.
\begin{equation}\label{eqn:cauchy-sol}
  f(x) = cx. \tag{$\clubsuit$}
\end{equation}
We gradually extend the domain for which we know this equation.
First, substitution $(x = x, y = -x)$ into the original equation gives $f(x) = -f(-x)$ which extends~\eqref{eqn:cauchy-sol} to all integers $x$.
Let $x$ be an arbitrary number satisfying~\eqref{eqn:cauchy-sol}, and $n$ be a positive integer. Then
\[ cx = f(x) = f\left(n \frac xn\right) = n f\left(\frac xn\right) \]
leading to
\[c \frac xn = f\left(\frac xn\right).\]
Therefore,~\eqref{eqn:cauchy-sol} holds for any rational number $x$.

Finally, we show that~\eqref{eqn:cauchy-sol} must be satisfied by every real number.
Suppose on the contrary that for some $x$, we have for example $f(x) < cx$, or $f(x) > cx$. We discuss here the first option, the other one is analogous.
If $f(x) < cx$, there is a rational number $q$ such that $\frac {f(x)}c < q < x$
(note $c > 0$ since $f$ is assumed to be increasing). Then $q < x$ but $f(x) < qc = f(q)$ contradicting that $f$ is increasing.

This proof goes beyond the abstraction level of SMT solvers.
Notably, it uses induction, which SMT solvers can
do~\cite{hozzova-cade21,hajdu,reynolds-vmcai15} but the induction here is over a
subdomain (natural numbers) vs.\ reals, where it cannot be applied directly.

\section{Template-and-QE}\label{sec:solving}
    
One could see the problem as quantifier elimination in the theory of
uninterpreted functions, accompanied by the theory required for describing the
problem itself --- in our case, the theory of reals or rationals. Due to the
undecidability of such a problem, quantifier elimination algorithms cannot be
used directly. A possible approach would be to \emph{synthesize} one function
adhering to the specification and then strengthen the specification so that the
individual solution is not admitted any more. Then, one would have to repeat
this process with the hope that there are finitely many solutions, or, observe
the solutions and generalize them into a more compact description. All these
steps are highly nontrivial. Instead, we propose an approach that tries a fixed
template and then performs quantifier elimination: \emph{template-and-QE},
which comprises the following subtasks.

\begin{enumerate}
    \item \emph{Identify} a template for the solution, e.g.\ $f(x)\triangleq ax+b$.
    \item\label{prove} \emph{Prove} that all solutions must fall into this template (\emph{template verification})
    \item Perform \emph{quantifier elimination} over input variables of the function(s).
\end{enumerate}

Note that in the case of reals, all tasks are computable, except for task~\ref{prove}.
The remainder of this section elaborates on the individual steps.

\subsection{Template Verification}\label{sec:template}

We run \cvc~\cite{cvc5}, \zthree~\cite{z3}, Vampire~\cite{vampire} and
Waldmeister~\cite{Hillenbrand2002} to attempt to prove all solutions
to a problem necessarily fit a certain template.
Since Waldmeister does not support theories directly, it needs special treatment
and we elaborate on this below. The other solvers
support natively the SMT2 syntax~\cite{smt2} enabling us to use the
combination of quantified non-linear reals with uninterpreted functions.\footnote{%
  The standard does not list \texttt{UFNRA} as one of the logics,
therefore the files indicate \texttt{AUFNIRA} as the closest superset.}

We consider the following set of templates for solutions: \emph{constant},
\emph{(monomial)\hspace{1pt}linear}, \emph{(monomial)\hspace{1pt}quadratic}. All
these are subclasses of the quadratic form but we consider these to simplify the
task for the solvers---it is conceivable that it is easier to prove that the
function must be constant than to prove that it must be arbitrary quadratic.
Since the set of templates is small, we always try all of them.

In order to ask an SMT solver if all solutions must be linear,
we add the assertion that the function is not linear.
An obvious way to state that $f$ is linear is via
the formula $\exists a b:\R.\forall x:\R.f(x) = ax + b$.
We call the problem resulting from including the properties
of the original problem along with the negation of
$\exists a b:\R.\forall x:\R.f(x) = ax + b$
the \emph{first variant of the linear template verification}.
We had more success using a different test for linearity.
A function $f$ is linear if and only if $\forall x:\R.f(x) = (f(1) - f(0))x + f(0)$.
Note that this formulation avoids the use of existential quantifiers.
We call the problem resulting from including the properties of the original problem along
with the negation of $\forall x:\R.f(x) = (f(1) - f(0))x + f(0)$
\emph{the second variant of the linear template verification}.
If any SMT solver can determine either variant is unsatisfiable, then
we know all solutions must be linear.

We likewise have two variants for each of the other templates.
The formulations for the first variants simply use existential quantification
for the coefficients.
For the second variants, we use the following properties:
\begin{itemize}
\item \emph{constant}: $\forall x:\R. f(x) = f(0)$
\item \emph{monomial linear}: $\forall x:\R. f(x) = f(1)x$
\item \emph{linear}: $\forall x:\R. f(x) = (f(1)-f(0))x + f(0)$
\item \emph{monomial quadratic}: $\forall x:\R. f(x) = f(1)x^2$
\item \emph{quadratic}: $\forall x:\R. 2f(x) = ((f(1) + f(-1)) - 2f(0)) x^2 + (f(1) - f(-1))x + 2f(0)$
\end{itemize}

\subsubsection{Using Waldmeister}\label{sec:wald}
In addition to using SMT solvers, we also used Waldmeister~\cite{Hillenbrand2002}.
Waldmeister is an automated theorem prover specializing in first-order unit equality.
Unlike SMT solvers, Waldmeister has no information about integers or reals.
Consequently, in order to ask Waldmeister if all functions $f$ satisfying
some properties must be contained within the class given by a template,
we first declare a sort $R$ along with constants $0,1:R$ and operations
$+$, $-$, $\cdot$ satisfying properties of a commutative ring with identity.
Each of these properties is given by a unit equation.
We purposely use rings instead of fields (although $\R$ is a field).
This allows us to ignore division since the side condition that the denominator
is not zero would result in a clause that is not a unit equation.
We only generate such a problem for Waldmeister when all the properties
stated for $f$ in the problem are unit equations, only quantify over reals (as opposed to integers),
and do not use division. Furthermore, we require that all specific real numbers
mentioned in the properties correspond to integers (e.g., 0.0, 1.0, -1.0, etc.).
These restrictions allow us to formulate a unit equation to give Waldmeister
corresponding to each property of $f$ given in the problem.

In addition to the assumed unit equations, Waldmeister expects a unit equation as a goal
to prove. The goal varies based on the template we are targeting
and corresponds to the second variant of the problems for the SMT solver.
In each case we fix a constant $d$ of type $R$ (not mentioned in the assumptions).
For each template, the goal unit equation is as follows:
\begin{itemize}
\item \emph{constant}: $f(d) = f(0)$
\item \emph{monomial linear}: $f(d) = f(1)d$
\item \emph{linear}: $f(d) = (f(1)-f(0))d + f(0)$
\item \emph{monomial quadratic}: $f(d) = f(1)d^2$
\item \emph{quadratic}: $2f(d) = ((f(1) + f(-1)) - 2f(0)) d^2 + (f(1) - f(-1))d + 2f(0)$
\end{itemize}
Waldmeister uses Knuth-Bendix style completion~\cite{completion} on the assumed unit equations
finding a proof of the goal when both sides of the goal rewrite to a common term.


For 8 of the problems, Waldmeister can prove all solutions are within the expected class.
Two of these problems were not covered by SMT solvers: U25 and U87.
In both of these problems, the least template class is linear monomials.
We briefly discuss these two problems.

The problem U25 asks for all real functions $f$ satisfying
$f(x f(x) + f(y)) = y + {f(x)}^2$.
Musil~\cite{musil} gives a proof that the only two solutions are $f(x) = x$ and $f(x) = -x$.
Waldmeister's goal is to prove $f(d) = f(1) d$
from the ring identities and the equation above.
It is difficult to directly compare the aforementioned human proof
to the proof found by Waldmeister. However, there are some interesting common intermediate
results.
Both proofs prove
$f$ is involutive (i.e., $f(f(x)) = x$), 
${f(x)}^2 = x^2$ 
and
$f(0) = 0$. 
On the other hand, there are several identities in each proof that do not appear in the other.
Musil's proof~\cite{musil} proves $f$ is surjective and uses this fact,
while the Waldmeister proof cannot directly represent the concept of surjectivity.

Waldmeister also proves all solutions to the problem U87 are linear monomials.
The problem U87 asks for all real functions $f$ satisfying
$f(x+y^2+z) = f(f(x)) + yf(y) + f(z)$.
The (only) two solutions are $f(x) = x$ and $f(x) = 0$.
In this case, there is no corresponding proof by Musil~\cite{musil} as a point of comparison.
A number of intermediate identities derived by Waldmeister stand out, e.g.,
$f(f(0)) = 0$ 
and
$f(f(x)) = f(x)$. 

\subsection{Quantifier Elimination}\label{sec:qe}

Quantifier elimination (QE) is a method to take a general formula in a theory and produce an equivalent quantifier-free
formula.
So for instance, in $\exists x\in\R.a<x<b$, the quantified variable $x$ is eliminated as $a<b$.
In many cases, quantifier elimination serves as a theoretical tool to show that
a theory is decidable but also has a long tradition of improvements at the
algorithmic level~\cite{cad,cooper72,FerranteRackoff-siamjc75,loosW93,Davenport,bradley-manna07,bjorner-janota-lpar15}.
In our work, QE is used to reveal the
particular solutions within a class given by a template.

We assume the original problem only has one uninterpreted function, $f$, and it is
a unary function from reals to reals.
If we want to determine all functions $f$ of the
form $f(x) = ax+b$, we can simply inline $f$ in the properties,
replacing each occurrence $f(t)$ by $at+b$.
The resulting problem has no uninterpreted functions and is usually in
the theory of the reals. (Exceptional cases are when the properties mention
integers as well as reals.) When the problem and template result in such
an inlined problem in the theory of reals, we can apply quantifier elimination.

We have made use of two implementations of quantifier elimination for the reals:
\zthree~\cite{z3} and \tarski~\cite{tarski}.
In \zthree, QE is invoked by applying the \texttt{qe} tactic, with the
\texttt{qe-nonlinear} parameter turned on.
The \tarski system is highly configurable but since QE is not a bottle-neck for
us, we use the \texttt{qepcad-qe}~\cite{CollinsH91} function in its default setting.
For each system, quantifier elimination should result in a quantifier-free formula involving
the uninterpreted constants (e.g., the coefficients $a$ and $b$ of the linear template
$ax+b$).
One might expect this formula to be in a solved form, e.g., $a=1 \lor a=0\land b=1$,
from which one can read off the precise class of linear functions satisfying the original
property.
However, this is generally not the case.
A separate postprocessing step attempts to convert the quantifier-free formula
into such a solved form. In practice the postprocessing does not always succeed.
In the case of \zthree, we also call the tactics simplify and propagate-values
to have \zthree simplify the formula before applying the postprocessing to attempt to find a solved form.

Problem U91 asks for all functions satisfying $f((x-y)^2) = {f(x)}^2 - 2xf(y) + y^2$.
Using the techniques of Section~\ref{sec:template}, we can prove all solutions $f$ must be linear.
After replacing $f(x)$ with $ax+b$ and calling \zthree to do quantifier elimination,
we obtain the quantifier-free formulas $a=1$ and $b=0\lor b=1$.
In general, \zthree represents the result of QE as a conjunction of several subformulas. The formula $a=1\land (b=0\lor b=1)$ is almost in solved form, and we can
easily obtain the two solutions $f(x)=x$ and $f(x)=x+1$ via postprocessing.
\tarski's quantifier elimination returns the more complicated formula
\[-1+a = 0 \land b\geq 0 \land -1+b\leq 0 \land (b=0 \lor -1+b = 0).\]
The postprocessor is also able to obtain the two solutions from this formula.

Sometimes there is a parameterized class of solutions.
Problem U3, for example, asks for all functions satisfying
$f(x+y) = f(x) + y$.
The techniques of Section~\ref{sec:template} determine all solutions are linear.
After inlining $f(x) = ax+b$, we call quantifier elimination.
Both \zthree and \tarski produce the formula $-1+a = 0$. From this the postprocessor can
determine $a=1$ and $b$ is unconstrained. This gives the class of solutions
$f(x) = x+b$ where $b$ is a real number. Indeed, this is the class of all solutions,
as desired.

\subsection{Postprocessor to Obtain Solved Forms}

We say a quantifier-free formula is
in \emph{solved form}\footnote{One may imagine loosening the concept but the general idea is to provide a specific solution as possible.} if it is a disjunctive normal form
where all the literals are of the form $c=v$
where $c$ is a coefficient of the template
and $v$ is a real number. We call a formula of the form $c=v$ an \emph{assignment atom}.
The postprocessor is given a list of formulas $\overline{\varphi}$ (thought of conjuctively)
and attempts to generate a solved form equivalent to $\varphi_1\land\cdots\land\varphi_n$.
As auxiliary values, a list $\overline{\iota}$ of equations of the form $s=t$ and
a list $\overline{\mu}$ of other literals
are generated, so that generally we are working with a triple
$(\overline{\varphi},\overline{\iota},\overline{\mu})$ of lists of formulas,
with $\overline{\iota}$ and $\overline{\mu}$ initially empty.
Suppose ${\overline{\varphi}}$ is nonempty, with $\varphi_1$ being the first formula on the list.
If $\varphi_1$ is a conjunction or a disjunction, we can simply make
recursive calls to the preprocessor to ask for solved forms of the decomposed formulas.
In the case of disjunction, we make two recursive calls (one for each disjunct) and
combine the results as a disjunction of the two solved forms (or fail if one call fails).
If $\varphi_1$ is an equation $s=t$, we add this to the equation list $\overline{\iota}$.
If $\varphi_1$ is of the form $s\leq t$, $s\geq t$, $s < t$ or $s > t$, then
we add this to the other list $\overline{\mu}$.
In every other case for $\varphi_1$, we simply fail to return a solved form.
We are finally left with the case where there are no remaining formulas in $\overline{\varphi}$.
The hope in this case is that $\overline{\iota}$ contains sufficient information
to obtain assignment atoms, and we only need to verify that the assignments
satisfy the other formulas from $\overline{\mu}$.
If the assignment atoms satisfy the other formulas, the conjunction of the assignment atoms
will be the solved form.
If the assignment atoms do not satisfy the other formulas, the empty disjunction $\bot$
will be the solved form.
It is also possible (e.g., if $c^2 = 1$ is in the list ${\overline{\iota}}$) 
a solved form with more than one disjunct.
At this point, the algorithm transforms a triple $(\overline{\iota},\overline{\mu},\overline{\alpha})$
where $\overline{\alpha}$ is the list of assignment atoms (initially empty).
We will also call $\overline{\alpha}$ an \emph{assignment}.

We say a term $t$ can be evaluated under an assignment $\overline{\alpha}$
if every constant occurring in $t$ is associated with a real number via $\overline{\alpha}$.
We say an atom $s=t$, $s\leq t$, $s\geq t$, $s<t$ or $s>t$ can be evaluated under an assignment $\overline{\alpha}$
if both terms $s$ and $t$ can be evaluated under an assignment $\overline{\alpha}$.
Clearly, if a term can be evaluated under an assignment $\overline{\alpha}$,
the term evaluates to a specific real number.
Likewise, if an atom can be evaluated under an assignment $\overline{\alpha}$,
the atom evaluates to true or false.

Assume ${\overline{\iota}}$ is nonempty and $s=t$ is the first equation on ${\overline{\iota}}$.
We consider the following cases (in order), dropping the equation from ${\overline{\iota}}$:
\begin{enumerate}
\item Assume both $s$ and $t$ can be evaluated under ${\overline{\alpha}}$.
  We drop the equation if they evaluate to the same reals and return $\bot$ otherwise.
\item Assume $s$ is a coefficient $c$ and $t$ can be evaluated to the real $v$.
  We add $c=v$ to ${\overline{\alpha}}$.
\item Assume $t$ can be evaluated under $\overline{\alpha}$
  and $s$ is one of a number of special forms which make use of an unassigned coefficient $c$
  and all other subterms $s_i$ that can be evaluated under $\overline{\alpha}$.
  Examples of these special forms are $s_1 + c$, $s_1+c s_2$ and $c^2$.
  In each case we solve for values $v$ of $c$ and add $c=v$ to $\overline{\alpha}$
  (or fail).
  If there are no solutions, we return $\bot$.
  If there are multiple solutions, we recursively call with both assignments 
  and return the disjunction of the resulting solved forms.
\item Assume the first equation $s=t$ is of the form $c+ s_1 d = t$
  and there is a second equation $s'=t'$ on the list ${\overline{\iota}}$
  of the form $c + d = t'$
  where $c$ and $d$ are unassigned coefficients and $s_1$, $t$ and $t'$ can
  all be evaluated under ${\overline{\alpha}}$.
  Then we solve these two linear equations for $c$ and $d$ (if possible)
  and add these to the assignment ${\overline{\alpha}}$, dropping
  both equations from ${\overline{\iota}}$.
\end{enumerate}

We finally assume ${\overline{\iota}}$ is empty.
If some formula in $\overline{\mu}$ cannot be evaluated via the assignment ${\overline{\alpha}}$,
then we fail to return a solved form.
Assume every $\overline{\mu}$ can be evaluated via the assignment ${\overline{\alpha}}$.
If every $\overline{\mu}$ evaluates to true, then we return the conjunction of the
assignment atoms in ${\overline{\alpha}}$ as the solved form.
Otherwise, we return $\bot$ as the solved form.

The postprocessor included sufficiently many cases to obtain a solved
form from either the output of \zthree or \tarski (or preferably both)
in each of the problems we considered.
There are many cases in which it will fail to compute a solved form.
Easy examples are $c^3 = 1$ and $2 c + d = 2 \land c + d = 1$.

\subsubsection{Lazy Verification}\label{sec:lazy}

We now briefly consider an alternative to the approach outlined above.
When considering a template, we do not need to verify that all solutions are in
the template class in advance.
Instead, we could use QE to find a
class of solutions for a fixed template and then prove there are no other
solutions. The advantage of this approach is illustrated by the following
example.
Problem C1 asks for all functions $f$ such that the following holds
\[ f(x+y)+2f(x-y)-4f(x) + xf(y) = 3y^2 - x^2 - 2xy + xy^2.\]
The only solution is $f(x) = x^2$.

In this case, the solvers listed in Section~\ref{sec:template} are unable
to prove that all solutions are a monomial quadratic (or that they must be quadratic).
However, we can still apply quantifier elimination to find all monomial quadratic solutions.
For this example, \zthree timed out even when given 10 minutes.
On the other hand, \tarski returned the formula $-1+a = 0$ (yielding the solution $a=1$, i.e., $x^2$)
in less than a second.
Since we were unable to prove all solutions must belong to the monomial quadratic template class,
it is still possible there are more solutions outside the monomial quadratic template class.
However, now that we have the specific solution $f(x)=x^2$ we can
ask an SMT solver to prove this is the only solution by assuming the equation
and also the negation of $\forall x:\R.f(x)=x^2$.
The solver \cvc is easily able to show this is unsatisfiable, so we know that $f(x)=x^2$ is the only solution.

\section{Experiments}\label{sec:experiments}

\begin{table}[t]
    \centering
    \caption{Result summary. Problems without any result are omitted.}%
    \subfloat[%
    \label{tab:gencov}
    Solution Template.
        Proven templates are denoted by~$\checkmark$;
        disproven template by $\times$;
        dash when no solver provided an answer.
    ]{%
    \begin{tabular}{lccccc}
\toprule
Problem & $c$ & $ax$ & $ax+b$ & $ax^2$ & $ax^2+bx+c$ \\ \midrule
\textbf{C2} & $\times$ & $\times$ & $\checkmark$ & $\times$ & $\checkmark$ \\
C6 & $\times$ & $\times$ & - & $\times$ & - \\
\textbf{C9a} & - & - & - & $\checkmark$ & - \\
\textbf{C10} & $\times$ & $\checkmark$ & $\checkmark$ & $\times$ & $\checkmark$ \\
C12 & $\times$ & - & - & $\times$ & - \\
C13 & $\times$ & $\times$ & $\times$ & $\times$ & $\times$ \\
U2 & $\times$ & - & - & $\times$ & - \\
\textbf{U3} & $\times$ & $\times$ & $\checkmark$ & $\times$ & $\checkmark$ \\
\textbf{U5} & $\times$ & $\checkmark$ & $\checkmark$ & $\times$ & $\checkmark$ \\
U7 & $\times$ & $\times$ & - & $\times$ & - \\
\textbf{U9} & $\times$ & $\times$ & $\checkmark$ & $\times$ & $\checkmark$ \\
\textbf{U13} & $\times$ & $\checkmark$ & $\checkmark$ & $\times$ & $\checkmark$ \\
\textbf{U16} & $\times$ & $\times$ & $\checkmark$ & $\times$ & - \\
U20 & $\times$ & $\times$ & - & $\times$ & - \\
U23 & - & $\times$ & - & $\times$ & - \\
\textbf{U24} & $\checkmark$ & $\checkmark$ & $\checkmark$ & $\checkmark$ & $\checkmark$ \\
\textbf{U25} & $\times$ & $\checkmark$ & $\checkmark$ & $\times$ & - \\
U26 & $\times$ & - & - & $\times$ & - \\
U27 & $\times$ & $\times$ & $\times$ & $\times$ & $\times$ \\
U41 & - & $\times$ & - & $\times$ & - \\
U42 & $\times$ & $\times$ & - & $\times$ & - \\
U44 & $\times$ & - & - & - & - \\
U45 & $\times$ & - & - & $\times$ & - \\
U48 & $\times$ & - & - & $\times$ & - \\
U49 & $\times$ & - & - & $\times$ & - \\
U50 & $\times$ & - & - & $\times$ & - \\
U51 & $\times$ & $\times$ & - & $\times$ & - \\
U54 & $\times$ & $\times$ & $\times$ & $\times$ & $\times$ \\
U56 & $\times$ & $\times$ & - & $\times$ & - \\
U62 & $\times$ & - & - & $\times$ & - \\
U64 & $\times$ & - & - & $\times$ & - \\
U67 & $\times$ & $\times$ & $\times$ & $\times$ & $\times$ \\
U68 & $\times$ & - & - & $\times$ & - \\
\textbf{U71} & $\checkmark$ & $\checkmark$ & $\checkmark$ & $\checkmark$ & $\checkmark$ \\
U72 & - & $\times$ & - & $\times$ & - \\
\textbf{U87} & $\times$ & $\checkmark$ & $\checkmark$ & $\times$ & $\checkmark$ \\
U90 & $\times$ & $\times$ & $\times$ & $\times$ & - \\
\textbf{U91} & $\times$ & $\times$ & $\checkmark$ & $\times$ & $\checkmark$ \\
U92 & $\times$ & - & - & $\times$ & - \\
\bottomrule
\end{tabular}

  }
  \hspace{1cm}
    \subfloat[Prove/check]{%
    \label{tab:prove}
    \begin{tiny}
    \begin{tabular}{lcc}
\toprule
Problem & Prove & Check \\ \midrule
\textbf{C1} & $\checkmark$ & $\checkmark$ \\
\textbf{C2} & $\checkmark$ & $\checkmark$ \\
C3 & - & $\checkmark$ \\
C4 & - & $\checkmark$ \\
\textbf{C5} & $\checkmark$ & $\checkmark$ \\
C6 & - & $\checkmark$ \\
\textbf{C9} & $\checkmark$ & $\checkmark$ \\
\textbf{C9a} & $\checkmark$ & $\checkmark$ \\
\textbf{C10} & $\checkmark$ & $\checkmark$ \\
C12 & - & $\checkmark$ \\
C13 & - & $\checkmark$ \\
U2 & - & $\checkmark$ \\
\textbf{U3} & $\checkmark$ & $\checkmark$ \\
\textbf{U4} & $\checkmark$ & $\checkmark$ \\
\textbf{U5} & $\checkmark$ & $\checkmark$ \\
U6 & - & $\checkmark$ \\
U7 & - & $\checkmark$ \\
U8 & - & $\checkmark$ \\
\textbf{U9} & $\checkmark$ & $\checkmark$ \\
U10 & - & $\checkmark$ \\
\textbf{U11} & $\checkmark$ & $\checkmark$ \\
U12 & - & $\checkmark$ \\
\textbf{U13} & $\checkmark$ & $\checkmark$ \\
U14 & - & $\checkmark$ \\
\textbf{U16} & $\checkmark$ & $\checkmark$ \\
\textbf{U17} & $\checkmark$ & $\checkmark$ \\
U19 & - & $\checkmark$ \\
U20 & - & $\checkmark$ \\
U23 & - & $\checkmark$ \\
\textbf{U24} & $\checkmark$ & $\checkmark$ \\
U25 & - & $\checkmark$ \\
U26 & - & $\checkmark$ \\
U27 & - & $\checkmark$ \\
U39 & - & $\checkmark$ \\
U41 & - & $\checkmark$ \\
U42 & - & $\checkmark$ \\
U44 & - & $\checkmark$ \\
U45 & - & $\checkmark$ \\
U46 & - & $\checkmark$ \\
U48 & - & $\checkmark$ \\
U49 & - & $\checkmark$ \\
U50 & - & $\checkmark$ \\
U51 & - & $\checkmark$ \\
U53 & - & $\checkmark$ \\
U54 & - & $\checkmark$ \\
U56 & - & $\checkmark$ \\
\textbf{U57} & $\checkmark$ & $\checkmark$ \\
U61 & - & $\checkmark$ \\
U62 & - & $\checkmark$ \\
U64 & - & $\checkmark$ \\
U66 & - & $\checkmark$ \\
U67 & - & $\checkmark$ \\
U68 & - & $\checkmark$ \\
\textbf{U71} & $\checkmark$ & $\checkmark$ \\
U72 & - & $\checkmark$ \\
U75 & - & $\checkmark$ \\
U76 & - & $\checkmark$ \\
\textbf{U79} & $\checkmark$ & $\checkmark$ \\
\textbf{U81} & $\checkmark$ & $\checkmark$ \\
\textbf{U82} & $\checkmark$ & $\checkmark$ \\
U87 & - & $\checkmark$ \\
U89 & - & $\checkmark$ \\
U90 & - & $\checkmark$ \\
U91 & - & $\checkmark$ \\
U92 & - & $\checkmark$ \\
U93 & - & $\checkmark$ \\
\bottomrule
\end{tabular}

    \end{tiny}
  }
\end{table}

We report the results of the template-and-qe method (Section~\ref{sec:solving}) on
the benchmark presented in Section~\ref{sec:problems}. The lazy extension of our approach (Section~\ref{sec:lazy}) was left for future work.
The template verification
(Section~\ref{sec:template}) has proven to be the most difficult task.
Table~\ref{tab:gencov} summarizes the obtained results for template
verification. For all these, we were able to solve the QE task.
Thirteen instances were solved completely by the automated
method. Out of these thirteen, two can be traced  to a competition event.
Problem U24 comes from Baltic Way competition---see Section~\ref{sec:problems}
for ``human'' solution. Problem U71 comes from The Prague Seminar
(PraSe-18-6-1). In both cases, the only solution is a function that is
constantly 0. 

The verification tasks results often in satisfiable problems, namely if there
exists a solution outside of the proposed template (the $\times$ symbol in
Table~\ref{tab:gencov}). These satisfiable problems are nontrivial
because they involve creating a counterexample to the template. For this
purpose, we also ran the SyGuS approach of \cvc (option
\texttt{--sygus-inference}) and its linear model
builder~\cite{janota-synasc2023}. 

We also report on the verification of the
correctness of the handwritten solutions, which comprises two components:
\emph{prove} --- show that all possible solutions are covered by the suggested
solutions; \emph{check} --- check that all suggested solutions are indeed
solutions to the problem (see Section~\ref{sec:problems}).
Table~\ref{tab:prove} summarizes the obtained results.
Sixty-five of the provided handwritten solutions were successfully checked,
i.e., the individual solutions satisfy the original specifications. Twenty of
the handwritten solutions were shown to cover all the individual solutions.
Unsurprisingly, the proving task is harder than the checking task.
We remark that checking solutions for \texttt{U79} is trivial because there are
no individual solutions.
More detailed results can be found on the authors' web page~\cite{results}.

\section{Conclusions and Future Work}%
\label{sec:conc}

This paper joins the effort of rising to the challenge of making
computers as powerful as the golden medalists at the International Math
Olympics~(IMO).
Efforts such as AlphaGeometry~\cite{Trinh2024}, show that machine
learning models are useful for the task but at the same time,
further research shows that they benefit from existing symbolic
methods~\cite{sinha2024wus}. Here we show that symbolic
methods are indeed already powerful in solving a highly nontrivial
task of finding \emph{all} functions fulfilling a certain specification.
Besides the task leading to undecidable questions, its difficulty also lies in
the fact it is \emph{not} a decision problem but the response is a description
of a class of mathematical objects.

The method we employ is anchored in templates, which is a well-known technique
in function and program synthesis~\cite{srivastava13}. The templates we
consider are rather simple and they could be made more powerful by adding
conditionals (if-then-else). However, the templates need to be kept simple if
we wish to be able to apply quantifier elimination.

An alternative to templates would be to attempt extending different synthesis
approaches, e.g., such as those that are realized in saturation-based
solvers~\cite{hozzova-cade23,hozzova-ijcar24} or inside SMT
solvers~\cite{reynolds-fmsd19,abate-jar23,barbosa-fmcad19}
or inductive logic programming~\cite{Cerna2024}.
The challenge here would be how to come up with all the solutions.

\begin{acknowledgments}
The results were supported by the Ministry of Education, Youth and Sports within
the dedicated program ERC~CZ under the project \emph{POSTMAN} no.~LL1902 and
co-funded by the European Union under the project \emph{ROBOPROX}
(reg.~no.~CZ.02.01.01/00/22\_008/0004590). C.\ Brown was supported by
\emph{CORESENSE}: the European Union’s Horizon Europe research and innovation
programme under grant agreement no.~101070254 CORESENSE\@.
This article is part of the \emph{RICAIP} project that has received funding
from the European Union's Horizon~2020 research and innovation programme under
grant agreement No~857306.
\end{acknowledgments}

\bibliography{synth,refs,qe}
\end{document}